\documentclass[conference]{IEEEtran}
\usepackage{setspace}
\usepackage[dvipsnames]{xcolor}

\usepackage{xspace} %abreviazioni nella scrittura 

\usepackage{amsmath}
\usepackage{amsthm}
\usepackage{amsfonts}
\usepackage{amssymb}
\usepackage{mathtools}
\usepackage{color}
\usepackage{tikz}
\usetikzlibrary{calc}
\usetikzlibrary{positioning}
\usetikzlibrary{shadows}
\usetikzlibrary{fit}
\usetikzlibrary{backgrounds}
\usetikzlibrary{arrows}
\usetikzlibrary{shapes}
\usetikzlibrary{matrix}
\usetikzlibrary{patterns}
\usetikzlibrary{automata}

\tikzstyle{every state}=[draw=black,minimum size=18,inner sep=0,fill=white!65,circle,text=black]

\usepackage{todonotes}

\newtheorem{mydef}{Definition}
\newtheorem{mythm}{Theorem}

\usepackage{multirow}
\IEEEoverridecommandlockouts
\IEEEpubid{\makebox[\columnwidth]{978-1-5090-5788-7/17/\$31.00~\copyright~2017
		IEEE \hfill}
	\hspace{\columnsep}\makebox[\columnwidth]{}}

\begin{document}

%% Paper title

\title{Automated Fixing of Access Policy Implementation in Industrial Networked Systems}

%% Author names and affiliations
\author{
	\IEEEauthorblockN{
										Manuel Cheminod, Luca Durante, Lucia Seno, Fulvio Valenza, Adriano Valenzano
	}
	\IEEEauthorblockA{
									National Research Council of Italy (CNR--IEIIT), 
									Corso Duca degli Abruzzi 24, I-10129 Torino, Italy\\
									Emails: \{manuel.cheminod, luca.durante, lucia.seno, fulvio.valenza, adriano.valenzano\}@ieiit.cnr.it
									%%
									%%%%%Alternativo
									%%
									%%CNR-IEIIT, c.so Duca degli Abruzzi 24, Torino I-10129, Italy, Corso Duca degli Abruzzi 24, I-10129 Torino, Italy\\ 
									%%Email: \{\}@ieiit.cnr.it
	}
\thanks{ \hspace{-0.4cm} 978-1-5090-5788-7/17/\$31.00~\copyright~2017 IEEE}
}

%% Title
\maketitle
\thispagestyle{plain}

%% Abstract
\begin{abstract}
Access control (AC) is the core of every architectural solution for information security. Indeed, no effective protection scheme can abstract from the careful design of access control policies, and infrastructures underlying modern Industrial Networked Systems (INSs) are not exceptions from this point of view. This paper presents a comprehensive framework for INS access control. The proposed approach enables the description of both positive and negative AC policies, by applying the Role Based Access Control (RBAC) paradigm to typical INS implementations, while taking into account different levels of abstraction. Suitable techniques are adopted  to check whether or not policies are correctly implemented in the system (verification). When conflicts are detected,    possible (re)assignments of credentials to the system users are automatically computed, that can be adopted to correct anomalies (conflict resolution). 
\end{abstract}

\IEEEpeerreviewmaketitle

\section{Introduction}
\label{sec:introduction}
\newcommand{\ldid}[1]{\textcolor{blue}{#1}}

Protecting industrial networked systems (INSs) against cyber-threats is a recognized crucial task. In fact, because of INS cyber-physical nature,  security and safety are strictly interdependent, so that a security breach can cause severe damages not only to assets but also to people and the environment. Despite awareness is constantly rising, security struggles to become common practice in the design, deployment and operation of INSs, and the demand for comprehensive solutions has still to find satisfactory answers \cite{Cheminod_TII13}.

This work deals with access control (AC) in INSs. Access control is known as a main architectural element for the security of every IT systems. In particular, we are interested in the implementation of policies that are designed to regulate accesses to system resources, so as to prevent unwanted interactions of unauthorized users with the system itself.

The contribution of this paper builds on some promising results obtained with the  innovative model and techniques for the automated analysis of AC policies introduced in \cite{CSI_2015, Cheminod_et_al_TII_2015}. The solution presented there was able to capture both the high level definition of AC policies and their low level implementation details in real INSs. Typical examples of such kind of details are h/w and s/w components, physical locations, device interconnections, shared resources and transactions, users' credentials and so on. 
That approach is extended here by introducing suitable techniques for fixing errors (anomalies) in the policy implementation, when they are discovered by the analysis process. 
The abstract definition of AC policies relies on the well-known Role Based Access Control (RBAC) formalism \cite{Sandhu_IEEEC96, ANSI_INCITS_359_2012} and, according to the terminology adopted in \cite{CSI_2015}, we call {\em Specification Model} $\mathcal{S}$ this kind of high-level  description. Similarly, the low-level implementation details, which are captured by means of the ad-hoc description language discussed in \cite{CSI_2015}, is called {\em Implementation Model} $\mathcal{I}$.

As explained in \cite{Cheminod_et_al_TII_2015} an automated s/w tool is able to process both $\mathcal{S}$ and $\mathcal{I}$ and compute, for each system user, the sets $I$ and $S$ of all actions allowed by the two models. Discrepancies between the high-level policies and the system implementation are then discovered by comparison of the two sets. In this way a confirmation/denial of policy correct implementation in the system is achieved.

This paper deals with a further important step, that is an automated technique to fix anomalies that are detected through the comparison above. This is obtained by searching for suitable (re)assignments of credentials, so that each user in the implementation model $\mathcal{I}$ is enabled to perform only those actions explicitly allowed by the specification model $\mathcal{S}$.

The paper is structured as follows:
Sect. \ref{sec:related_works} deals with some relevant works appeared in the literature, that focus on access control policy verification and anomaly resolution for general purpose networked systems. 
% Sect. \ref{sec:back} sheds some light on the concepts, algorithms and results of \cite{CSI_2015, Cheminod_et_al_TII_2015}, useful and needed here.
Sect. \ref{sec:specification_model} recalls some basic elements of the approach in \cite{CSI_2015, Cheminod_et_al_TII_2015}, that are needed to understand the remaining part of the paper. It also introduces extensions and changes to the specification  model to manage anomalies automatically. Sect. \ref{sec:hints_on_the_approach} presents the proposed resolution method, while in Sect. \ref{sec:clarifying_example} a simple example is described to offer the reader a flavour of our technique at work. Finally, Sect. \ref{sec:conclusion_future_works} concludes the paper.

\section{Related Works}
\label{sec:related_works}
Access control has received significant attention in the scientific literature. In particular, the RBAC framework~\cite{Sandhu_IEEEC96,ANSI_INCITS_359_2012}, which relies on the definition of roles (representing responsibilities) to which system users and permissions are assigned, has progressively gained popularity because of its easy and flexible policy management.

Most works dealing with access control, e.g.,~\cite{Jayaraman_CCS11,Sun_TDSC11, Hughes2008} and \cite{Basile2017}, only tackle policy management issues. In particular, they focus on the analysis of sets of abstract policies (verification of specific requirements,  detection of conflicts and suboptimal descriptions) and do not take into account their actual implementation in real systems. In~\cite{Jayaraman_CCS11} a policy analysis tool is described,  which makes use of a model checker, while in~\cite{Sun_TDSC11,Hughes2008,Basile2017} queries about access control policy properties are translated into Boolean satisfiability problems and solved using a SAT (SATisfiability) solver. Some works such as ~\cite{Koch2002,Hu2011} propose resolution strategies besides the identification of conflicts, however their approaches only involve the policy definition domain. 

Some papers, e.g.,~\cite{Cau_FMSD13,Hinrichs_CSF13}, take into account the problem of enforcing policies in real systems, and propose solutions which, however, assume the availability of  suitable enforcement mechanisms. In practice, they rely on sophisticated h/w and s/w support which, unavoidably, cannot be provided by actual INSs, as they are typically characterized by limited computational and communication resources and strict real-time performance requirements.

Few works, e.g.,~\cite{skybox,Nicol_SP08,Nicol_HICSS10,NP-View,Okhravi_SafeConfig09,Le_2015_SACMAT}, deal with the implementation of AC policies in general-purpose systems without making assumptions about their enforcement. In particular, \cite{skybox} focuses on networked systems consisting of firewalls and other kinds of traffic control components and, as such, the presented approach is unsuitable for INSs which, typically, also include other types of devices. The NP-View tool described in~\cite{Nicol_SP08,Nicol_HICSS10,NP-View} allows to check global policy implementations by processing information about the network topology and configuration. The tool, however, can only be used for systems consisting of nodes running SE-Linux which is typically not the case of INSs. The approach proposed in~\cite{Okhravi_SafeConfig09} shares some similarities with~\cite{CSI_2015,Cheminod_et_al_TII_2015} as it is intended to compute ``who can do what on what'' but, as for NP-View, it needs all nodes to run SE-Linux and does not apply to INSs. Finally, \cite{Le_2015_SACMAT} deals with the automatic description of resource access information but, unfortunately, the solution only concerns web applications.

\section{Extended Formal Model}
\label{sec:specification_model}
The twofold model described in \cite{CSI_2015} and consisting of both $\mathcal{S}$ and  $\mathcal{I}$, enables the computation and comparison of two finite sets of actions $S$ and $I$ for the purpose of checking the correctness of policy implementation.
  
Elements in $S$ and $I$ are triples (user, operation, object), i.e. $(u, \pi, \omega) \mid u \in Users, \pi \in Operations, \omega  \in Objects$, meaning that user $u$ is allowed to do operation $\pi$ on object $\omega$. However, to find fixes automatically, ambiguities have to be eliminated. Roughly speaking, the abstract (RBAC-based) description should specify both actions that are enabled in the system and actions that must be forbidden. Actions not included in the explicit specification can then be treated as ``don't care'' situations and possibly leveraged to select a credential assignment which satisfies the policies in   $\mathcal{S}$.

The new extended specification model adopted in this paper allows designers to define two disjoint sets $S^{+}, S^{-}\subseteq Users \times Operations \times Objects$ describing, respectively, operations $\pi \in Operations$ allowed and forbidden (i.e. \textit{allowed} and \textit{denied} permissions) on system objects $\omega \in Objects$ by each user $u \in Users$.  Triples in $Users \times Operations \times Objects$  belonging to neither $S^{+}$ nor  $S^{-}$ concern permissions not relevant to access control, that is whose possible assignment to the user is not significant (\textit{don't care} permissions). Clearly $S^{+} \cap S^{-} =\emptyset$. 

The extended specification model $\mathcal{S}$ is still based on the RBAC paradigm \cite{ANSI_INCITS_359_2012, Sandhu_IEEEC96}, where the assignment of allowed and denied permissions to users is obtained through the definition of a suitable set of roles $Roles$ representing collections of responsibilities. In practice, each role $r \in Roles$ is assigned those users sharing that role and two sets of allowed/denied permissions through the definition of the following functions\footnote{In this paper $2^W$ is the power set of set $W$, i.e., the set of all possible subsets of $W$.} 
\begin{eqnarray}
usr\_asg&:& Roles \rightarrow 2^{Users}\\
prm\_asg^{+}&:& Roles \rightarrow 2^{Operations \times Objects}\\
prm\_asg^{-}&:& Roles \rightarrow 2^{Operations \times Objects}
\end{eqnarray}

Note that a user may be assigned many roles and different roles may be associated the permission. Moreover, since $prm\_asg^{+}(r) \subseteq S^{+}$ and $prm\_asg^{-}(r) \subseteq S^{-}$ $\forall r \in Roles$, clearly $prm\_asg^{+}(r) \cap prm\_asg^{-}(r)=\emptyset$.

When hierarchical RBAC is adopted and a hierarchical relation ($\prec$) is defined between elements of $Roles$, the assignments of users, allowed and denied permissions to roles is obtained as
\begin{eqnarray}
\overline{usr\_asg}(r)& ::=&\bigcup_{r^{\prime}\succeq r}usr\_asg(r^{\prime})\label{u_as}\\
\overline{prm\_asg}^{+}(r)& ::= &\bigcup_{r^{\prime}\preceq r}prm\_asg^{+}(r^{\prime})\label{p_asP}\\
\overline{prm\_asg}^{-}(r)& ::= &\bigcup_{r^{\prime}\succeq r}prm\_asg^{-}(r^{\prime})\label{p_asM}
\end{eqnarray}
Basically, users having role $r$ also inherit roles $r'\prec r$, and each role $r$, in addition to its explicitly allowed permissions (i.e., $prm\_asg^{+}(r))$ also inherits allowed permissions of roles $r'\prec r$. Denied permission inheritance propagates in the opposite direction in the role hierarchy.

Sets $S^{+}$ and $S^{-}$ can be easily computed by associating each user with her/his allowed and denied permissions through roles as follows (symbol $\bullet$ stands for either $+$ or $-$)
\begin{eqnarray}\label{eq:SpecSetDef}
S^{\bullet} \!\!\!\!\!\! &::=& \!\!\!\!\!\! \left \{(u, \pi, \omega ) \in Users \times Operations \times Objects \mid \right.\\ \nonumber
&&\left. \!\!\!\!\!\!\exists \ r \in Roles \mid u\in \overline{usr\_asg}(r), (\pi,\omega) \in \overline{prm\_asg}^{\bullet}(r)\right \}
\end{eqnarray}
When set in (\ref{eq:SpecSetDef}) is computed for given user $u^{\star} \in Users $, we use notation
\begin{eqnarray}
S^{\bullet}_{u^{\star}} &::=& \left \{(u, \pi, \omega) \in S^{\bullet}\mid u=u^{\star}\right\}\label{eq:SecSetDefUser} \\
\tilde{S}^{\bullet}_{u^{\star}} &::=& \left\{ (\pi,\omega) \mid (u^{\star}, \pi,\omega) \in S^{\bullet}_{u^{\star}}\right\} \label{eq:SecSetDefUserC}
\end{eqnarray}
where in (\ref{eq:SecSetDefUserC}) the leftmost component of the triple has been removed, being anyway the link to $u^{\star}$ kept by the subscript of $\tilde{S}^{\bullet}$.
%To concisely say that a specification model $\mathcal{S}$ can give both a set of allowed and denied permissions by following the method sketched above, let's write:
%\begin{equation*}
%\mathcal{S} \models (\bar{S}^{+}, \bar{S}^{-})
%\end{equation*}

With respect to \cite{CSI_2015}, no extension is needed for the implementation model $\mathcal{I}$. Of course, the computation of $I$ still remains a bit tricky and the reader may refer to  \cite{Cheminod_et_al_TII_2015} for details. For the purpose of this paper it is enough remembering that, starting from a detailed description of the real system elements (i.e. devices, rooms, services, configurations, network links, user credentials and so on), a suitable set of inference rules (describing the possible interactions of a generic user with the system) and the users' initial states, a suitable automaton $A_u$ can be built for each user $u$ ($u\in Users$)  describing all possible sequences of actions s/he can carry out on the system. Automaton edge labels $(\pi, \omega)$ augmented with the user identifiers $u$ are triples $(u, \pi, \omega)$  of set $I$. 

To decrease the computation complexity and prevent the typical state explosion problem,  \cite{Cheminod_et_al_TII_2015} showed how to compute the automata in an optimized way. In particular, the automaton $A_u$ for user $u$ is obtained as the parallel composition of $A^{r}_u$ and $A^{L,min}_u$ describing, respectively, the dynamic of the user moving in the system and its actual interaction with the system resources.  $A^{r}_u || A^{L,min}_u$ has, in general, a lower number of states than $A_u$, thus allowing optimization of storage resources and  computation. With a slight abuse of notation,  we use $A_u$ to mean $A^{r}_u || A^{L,min}_u$ as well.

Given the specification and implementation models $\mathcal{S}$ and  $\mathcal{I}$, respectively enabling the evaluation of sets $S^{+}$, $S^{-}$, and $I$, we use symbols $\bar{S}^{+}$ and $\bar{S}^{-}$ to mean the following sets
\begin{eqnarray}
\bar{S}^{+} ::= S^{+} \setminus I \label{eq:Sp} \\
\bar{S}^{-} ::= S^{-}\cap I \label{eq:Sm}
\end{eqnarray}
$\bar{S}^{+}$ consists of those triples representing actions allowed to users $u \in Users$, according to $\mathcal{S}$ which are actually not enabled by the system implementation $\mathcal{I}$, whereas $\bar{S}^{-}$ collects triples $(u, \pi, \omega)$ describing permissions assigned to users $u \in Users$ by $\mathcal{I}$, but forbidden to the same users by policies in $\mathcal{S}$. According to this description, we can say that
% the following relations hold true 
%
\begin{mydef}
Given a specification model $\mathcal{S}$ and an implementation model $\mathcal{I}$ respectively leading to sets $S^{+}$, $S^{-}$, and $I$ (and, consequently to $\bar{S}^{+}$ and $\bar{S}^{-}$), the system correctly implements the policies if and only if 
\begin{equation}\label{eq:Spm0}
\bar{S}^{+} = \emptyset \ \wedge \ \bar{S}^{-} = \emptyset 
\end{equation}
\end{mydef}
Whenever (\ref{eq:Spm0}) does not hold true, some anomalies are present in the policy implementation.
In particular, $\bar{S}^{+} \neq \emptyset$ means that some allowed permission of $\mathcal{S}$ is not implemented in $\mathcal{I}$, whereas when $\bar{S}^{-} \neq \emptyset$ some forbidden action of $\mathcal{S}$ is allowed in $\mathcal{I}$.
%Relation $(1)$ essentially states that any allowed permission (i.e., any element of set $S^{+}$) has to be actually permitted by the current system implementation (needs to belong to $I$), while relation $(2)$ states that none of the denied permissions (i.e., none of the elements of set $S^{-}$) should be actually performed by the user on the system (i.e., should be indeed also part of $I$).
We refer to the process of checking for the presence of anomalies as policy implementation analysis or verification.

We can now focus on the automatic correction of anomalies. When a system modeled by $\mathcal{I}$ does not correctly implement the policies modelled by $\mathcal{S}$ (i.e., either $\bar{S}^{+}$ or $\bar{S}^{-}$ are non-empty), the problem of anomaly resolution is that of finding suitable changes in the system implementation so that the detected anomalies are removed (i.e., modifying $\mathcal{I}$ such that (\ref{eq:Spm0}) holds). Note that not all modifications of $\mathcal{I}$ are admissible as the functionality of the system (currently not explicitly described by model $\mathcal{I}$) needs to be preserved.

In the following we concentrate on solutions based on changes in the assignments of credentials to users, i.e., restricting the space of possible solutions by leaving the system topology and device configurations unchanged. Since the considered modifications can only affect user initial states and credentials, this is clearly a first, preliminary step in the resolution process. The main advantage is that the problem of preserving the system functionality can be ignored, nevertheless useful insights can be obtained about the current system implementation and its relation with the detected anomalies.

\section{Policy Verification and Correction}
\label{sec:hints_on_the_approach}
Informally, the approach for the automated correction of policy implementation is based on the construction of a new kind of automaton, which extends $A_{u}$  discussed in \cite{Cheminod_et_al_TII_2015}  with the following characteristics:
\begin{itemize}
\item the behaviour of a (fictious) {\em super}-user is described. This super-user is assigned all credentials available in the system. 
\item The automaton edge labels also include the credentials enabling state transitions besides the conventional pairs $(\pi, \omega)$.
\end{itemize}
A suitable visit of this automaton allows the computation of the set of all operations on system objects that have to be performed prior to a given operation on a specific object can be executed. In doing this,  the visiting algorithm keeps track of needed credentials that can be deduced from the automaton edge labels. In this way, comprehensive solutions can be searched, that assign each user a suitable set of credentials enabling her/him to perform only those operations authorized by the specification policies.

Formally, the super-user automaton $A$, generating language $\mathcal{L}(A)$, is defined as
\begin{equation}\label{eq:Acap}
A ::= (Q, \Sigma, \delta, q^0)
\end{equation}
where $Q$ and $q^{0}$ are respectively the set of states and the initial state of the super-user, with the same structure described in \cite{Cheminod_et_al_TII_2015}.
With respect to $A_{u}$ for conventional users,  $A$ deals with {\em extended} events each one also taking into account the credential $c$ which is required to enable the event $(\pi, \omega)$. Formally, the set of extended events is $\Sigma ::= Operations \times Objects \times \{C \ \cup\ \{\varepsilon \}\}$ and the automaton transition function is $\delta: Q \times \Sigma \rightarrow Q$.
This means that performing the same operation $\pi$ on object $\omega$ by owning different credentials leads to different transitions in $A$ (i.e., $(\pi,\omega, c_1)\neq (\pi,\omega,c_2)$). Moreover, notation $(\pi,\omega,\varepsilon)$  is used for labels where operation $\pi$ is performed on object $\omega$ without owning any specific credential.

Evening in case of the super-user case too, $A$ can then be efficiently computed as the parallel composition of two simpler automata  $A^{r}$ and $A^{L,min}$, that is  $A=A^{r} || A^{L,min}$.

Roughly speaking, $A$ is essentially a super-automaton with respect to $A_u$, which is able to describe all possible sequences of operations a hypothetical super-user, owning all available credentials, can perform on the system objects. As a consequence, $A$ provides an overview of how access to the system resources can be obtained by following different paths, i.e., performing different sequences of operations and, possibly, exploiting different sets of credentials (this somehow resembles what happens in attack graphs). 

Once the structure of $A$ has been determined, the main idea is to use it to compute, for each pair $(\pi,\omega)$ such that  $\exists \ q, q^{\prime} \in Q, c \in C \mid \delta(q,(\pi,\omega, c)) = q^{\prime}$, the sets of operations which, when performed in a specific order, actually enable the execution of $(\pi,\omega)$ (i.e., change the state of the user so that s/he is allowed to perform $(\pi, \omega)$), and, from these, derive the sets of credentials that enable the operation itself. 
%\Lucia{Spiegare che l'idea è poi trovare una funzione booleana che dipende dalle credenziali possedute dall'utente}

To this purpose let us introduce the following definitions.

\subsection{Enabling events}

\begin{figure}
	\centering
	\includegraphics[width=0.7\linewidth]{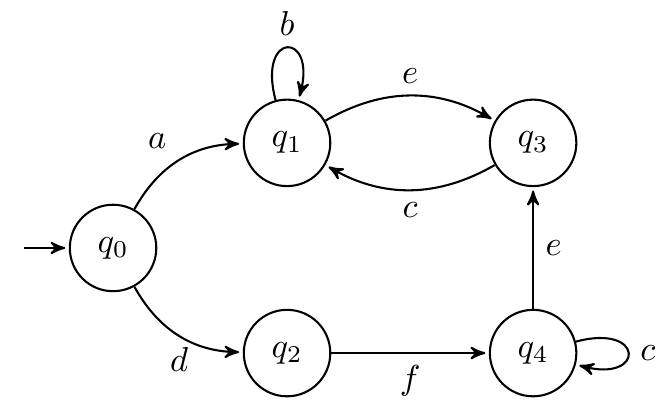}
	\caption{Example automaton $\overline{G}$}
	\label{fig:ae}
\end{figure}

\begin{mydef}
Given the set of events $\Sigma$ and a string $s \in \Sigma^{*}$,  function $T_{\Sigma}(s): \Sigma^{*} \rightarrow 2^{\Sigma}$ is defined as\footnote{Notation $W^{*}$ indicate the Kleene closure of set $W$, i.e. $W^{*}$ is the set of all strings of any length obtained concatenating elements of $W$.}:
\begin{equation*}
 T_{\Sigma}(s) ::= \left \{e \in \Sigma \mid \exists  \ t,v \in \Sigma^* : s=tev \right\}
\end{equation*}
\end{mydef}
Basically, given a sequence $s$ obtained as a concatenation of elements in $\Sigma$, function $T_{\Sigma}(s)$ returns the set of all events in $s$, i.e. it ``tokenizes'' $s$. As an example, if we consider string $s=abbe\in \mathcal{L}(\overline{G})$, where $\mathcal{L}(\overline{G})$ is the language generated by example automaton $\overline{G}=(Q, \Sigma, \delta, q_0)$ depicted in Fig. \ref{fig:ae}, $T_{\Sigma}(s)=\{a, b, e\}$.

\begin{mydef}
 Given $A=(Q, \Sigma, \delta, q_0)$, generating language $\mathcal{L}(A)$, event $e \in \Sigma$, and  set $V \in 2^{\Sigma \setminus \{e\}}$, function $L_{A} : 2^{\Sigma \setminus \{e\}} \times \{e\} \rightarrow \mathbb{B}$ is defined as:
 \begin{equation*}\begin{split}
 L_{A}(V,e) & ::= \\
 & \begin{array}{c}
                                   \exists s \in \Sigma^{*} \mid se \in \mathcal{L}(A), \ T_{\Sigma}(s) = V \\
                                  \wedge \\
                                 \nexists  V^{\prime} \subset V \mid \exists  s^{\prime} \in \Sigma^{*} \mid s^{\prime}e \in \mathcal{L}(A), \ T_{\Sigma}(s^{\prime}) = V^{\prime}
                     \end{array}
 \end{split}\end{equation*}
Given $A$, $V$, and $e$ defined as above, we say that $V$ is an enabling set of events for $e$ in $A$. $V \in 2^{\Sigma \setminus \{e\}}$ means $V \subseteq \Sigma \setminus \{e\}$, whereas $\mathbb{B}$ stands for the {\em boolean} domain.
\end{mydef}

Intuitively, given a set $V$ of events belonging to $\Sigma\setminus \{e\}$, and event $e \in \Sigma$, where $\Sigma$ is the set of events of a certain automaton $A$, $L_{A}(V,e)$ is {\em true} if and only if events of $V$ belong to some path in $A$ (not containing $e$), where $e$ is the next event, and there is not $V^{\prime} \subset V$ with the same property.
To clarify the notion of enabling set of events, let us refer again to Fig. \ref{fig:ae}. We consider event $e\in\Sigma$ and sets $V_1, V_2, V_3 \in \Sigma \setminus \{e\}$ defined as $V_1={\varepsilon}$, $V_2=\{a\}$ and $V_3=\{a,b\}$. By computing $L_{\overline{G}}(V_i,e)$ we obtain that while $V_2$ is an enabling set of events for $e$ in $G$, $V_1$ and $V_3$ are not as $e \notin\mathcal{L}(\overline{G})$ and even if $abe \in\mathcal{L}(\overline{G})$, $V_2\subset V_3$.

\begin{mydef}
 Given $A=(Q, \Sigma, \delta, q_0)$, and event $e \in \Sigma$, function $E_{A}: \Sigma \rightarrow 2^{2^{\Sigma \setminus \{e\}}}$ is
 \begin{equation}
 E_{A}(e)=\left \{ V \in 2^{\Sigma \setminus \{e\}}: L_{A}(V,e) \right \}
 \end{equation}
\end{mydef}

Definitions above state that, given a generic automaton describing a system behavior, any event $e$ in event set $\Sigma$ is characterized by a sets enabling set of events $E_{A}(e)$. Event $e\in \Sigma$ may occur (is enabled) if and only if all events in any of the sets in $E_{A}(e)$ have already occurred. Note that the  set of enabling events for $e\in \Sigma$ may be $E_{A}(e)=\{\{\varepsilon\}\}$ meaning that $e$ is enabled directly in the initial state of $A$ (i.e., it is not enabled by other events). Moreover, $E_{A}(e)=\{\{\varepsilon\}\}$ is not the same as $E_{A}(e)=\emptyset$, as the latter indicates that event $e\in\Sigma$ can never happen in $A$ (i.e., no transition in $A$ is associated to event $e$).

Referring again to the example automaton $\overline{G}$ and considering event $e\in \Sigma$, we obtain that $E_{\overline{G}}(e)=\{\{a\}, \{d,f\}\}$, meaning that, $e$  necessarily follows either event $a$ or both events $d$ and $f$ in  $\overline{G}$.
Several algorithms allow the computation of $E_{A}(e)$ given $A$  and $e\in\Sigma$.

\begin{mydef}
	Given an automaton $A=(Q, \Sigma, \delta, q_0)$, an event $\tilde{e} \in \Sigma$, and the function $E_{A}$, we define function $F_{A}(\overline{e})$ as 
\begin{equation}\label{eq:Fe_gen}
{F}_{A}(\overline{e}) ::=  \sum_{ V \in E_{A}(e)}\left(\prod_{e \in V}e\right)
\end{equation}
\end{mydef}

Referring again to the example automaton $\overline{G}$, and considering event $\overline{e}\in \Sigma$, we obtain that $F_{\overline{G}}(\overline{e})=a+d \cdot f$. In the following we will deal with automata as (\ref{eq:Acap}), and we will provide some refinement of the above definitions, in particular Def. 5 will be enhanced by Def. 8.

\subsection{Enabling function computation}
%\All{Forse, da un punto di vista strettamente formale, la cosa migliore sarebbe chiamare $\Sigma$ ed $e$ rispettivamente gli insiemi di eventi e gli eventi degli automi di [3] e [4], e qui usare una notazione estesa per indicare gli eventi estesi con la credenziale ed i relativi insiemi, ad esempio usando sempre il cappello che abbiamo gi\`a messo all'automa del superuser: $\hat{e}$ e $\widehat{\Sigma}$.}

%In automaton $\widehat{A}$ events are in the form $\hat{e}=(\pi,\omega, c)\in Operations \times Objects \times \{C \ \cup\ \varepsilon \}$. Given for each $\hat{e}=(\pi,\omega, c)$ appearing on the transitions of  $\widehat{A}$ the set  of sets $E_{G}(\hat{e})$, we would like to build
%a boolean function, which given the set of credentials own by a generic user, returns true if the user is able to perform $(\pi,\omega)$ and false otherwise. To this purpose, we provide the following auxiliary definitions

\begin{mydef}
Given the set of events $e = (\pi,\omega,c)\in\Sigma$, let us define
\begin{eqnarray*}
c(e) & ::= & c \\
\mathcal{C}(\Sigma) & ::=&  \left\{ c(e) \mid e \in \Sigma \right\} \\
\tilde{e} & ::= & (\pi, \omega) \mid e = (\pi, \omega, c) \\
\tilde{\Sigma} & ::= & \left\{ \tilde{e} \mid e \in \Sigma \right\}
\end{eqnarray*}
\end{mydef}
In practice, with little abuse of notation, $c(e)$ and  $\mathcal{C}(\Sigma)$ respectively return the credential $c$ of event $e$ and the set of the credentials of $\Sigma$, whereas the remaining  two functions drop the credential from the event representation(s). We call $\tilde{e}$ and $\tilde{\Sigma}$ respectively {\em reduced event} and {\em set of reduced events} and, in the following, $(\pi, \omega)$ is referred to as reduced event independently on whether it is obtained from some $\Sigma$.
%\Luca{Qui un esempietto \`e banale, ma forse utile}

\begin{mydef}
Given  $\Sigma$ and the reduced event $\tilde{e} = (\pi,\omega)$, $\Sigma_{\tilde{e}}$ is
\begin{equation*}
\Sigma_{\tilde{e}} ::= \left\{ (\pi, \omega, c) \in \Sigma \mid (\pi, \omega) = \tilde{e} \right\}
\end{equation*}
\end{mydef}

\begin{mydef}
Given $A$, its set of events $\Sigma$, the corresponding set of reduced events $\tilde{\Sigma}$ and the reduced event $\tilde{e}$, $\mathcal{F}_{A}(\tilde{e})$ is 
\begin{equation}\label{eq:Fe}
% \mathcal{F}(e) ::= \bigvee_{\forall \hat{e} \in \widehat{\Sigma}_{e}} \left( \bigvee_{\forall V \in E_{G}(\hat{e})}\left(\bigwedge_{\forall c \in \mathcal{C}(V \cup \left\{ c_{\hat{e}}\right\}}c\right)\right)
\mathcal{F}_{A}(\tilde{e}) ::= \sum_{e \in \Sigma_{\tilde{e}}} \left( \sum_{ V \in E_{A}(e)}\left(\prod_{c \in \mathcal{C}(V) \cup \left\{ c(e)\right\}}c\right)\right)
\end{equation}
\end{mydef}
$\mathcal{F}_{A}(\tilde{e})$ is used to build an expression containing symbols $c$ i.e. credentials identifiers, and operators $\cdot$ and $+$. In the following, such an expression is treated as a boolean function where symbols $c$ are boolean variables. From a practical point of view the meaning of $c$ here is twofold: $s$ is a credential identifier when (\ref{eq:Fe}) is constructed, but also a boolean variable when (\ref{eq:Fe}) is computed.
%{\em where $\prod$ stands for the symbol of} logical {\em product} AND, {\em and $\sum$ stands for the symbol of } logical {\em sum} OR, and the credentials {\em id}s are boolean variables.

\begin{mydef}
Given $A$ and the reduced event $\tilde{e}$ leading to function $\mathcal{F}_{A}(\tilde{e})$ (\ref{eq:Fe}), user $u$ and set of credentials $C$ belonging to $u$'s initial state, 
$\mathcal{F}_{A}(\tilde{e})|_{u}$ is $\mathcal{F}_{A}(\tilde{e})$ where each boolean variable $c$ is set to $1$ if and only if $c \in C$, and $0$ otherwise.
\end{mydef}

Roughly speaking, $\mathcal{F}_{A}(\tilde{e})|_{u}$ is the result of the computation of $\mathcal{F}_{A}(\tilde{e})$ with each variable bound to $1$ if and only if user $u$ owns a credential with the same name, and bound to $0$ otherwise.

 %\label{eq:SecSetDefUser}
 
\begin{mythm}
Given a system modeled by implementation model $\mathcal{I}$ (and automaton $A$ derived from $\mathcal{I}$), some access control policies modeled by specification model $\mathcal{S}$ (leading to specification sets $S^{+}$, $S^{-}$), any correct assignment of user credentials  (i.e., any user credential assignment preventing implementation anomalies) is such that for each $u \in Users$ the following holds true

\begin{equation}\label{sol}
\left(\bigwedge_{\tilde{e} \in \tilde{S}^{+}_u} \mathcal{F}_{A}(\tilde{e})|_{u}\right) \wedge \left(\bigwedge_{\tilde{e} \in \tilde{S}^{-}_u} \overline{\mathcal{F}_{A}(\tilde{e})|_{u}}\right)=1
\end{equation}

where sets $\tilde{S}^{+}_u$ and $\tilde{S}^{-}_u$ come from $\mathcal{S}$ through (\ref{eq:SecSetDefUserC}).
\end{mythm}
To make the reading lighter, the proof has been omitted here. 
Note that the set of equations (\ref{sol}) may have one solution, multiple solutions or no solution at all. A possible way to compute a solution (if any exists) is to use a SAT solver.

\section{Example}
\label{sec:clarifying_example}
\begin{figure}
	\centering
	\includegraphics[width=0.7\linewidth]{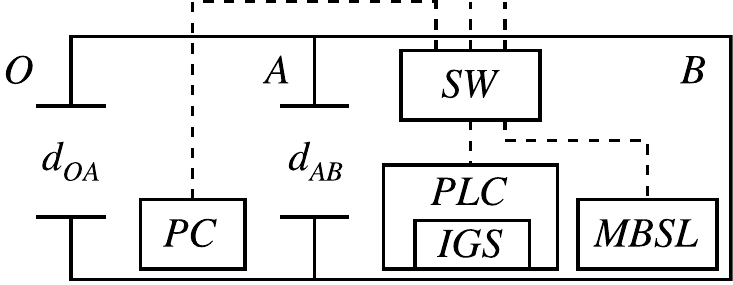}
	\caption{Room and network topology.}
	\label{fig:cabinet}
\end{figure}

\subsection{System description}
\begin{table}[t]
		\caption{Specification model $\mathcal{S}$}\label{tab:policies}
	\resizebox{\columnwidth}{!}
	{
		\centering
		\renewcommand{\arraystretch}{1.3}
		\begin{tabular}{l|l|l} 
			\hline
			\multirow{2}{*}{$P_{o}$}  & $prm\_asg^+(P_o)$ 	& $\{(run, MBSL), (run, IGS)\}$ \\ \cline{2-3}
			& $prm\_asg^-(P_o)$ 	& $\{(admin, MBSL), (admin, IGS),  (admin, PLC)\}$  \\ \cline{2-3}
			& $usr\_asg(P_o)$ 	& $\{Tom\}$\\ \hline	
			\multirow{2}{*}{$P_{s}$}  & $prm\_asg^+(P_s)$ 	& $\{(run, MBSL), (run, IGS), (admin, MBSL),$ \\
			&&$ (admin, IGS), (admin, PLC) \}$\\ \cline{2-3}
			& $prm\_asg^-(P_s)$ 	& $\emptyset$\\ 	\cline{2-3}
			& $usr\_asg(P_s)$ 	& $\{Amy\}$\\ \hline
		\end{tabular}
		\renewcommand{\arraystretch}{1.0}
	}
\end{table}
\begin{table}[t]
	\centering
		\caption{Users' sets of assigned credentials $C_u$}\label{tab:credential}
	\resizebox{\columnwidth}{!}
	{
		\footnotesize 
		\centering
		\renewcommand{\arraystretch}{1.3}
		\begin{tabular}{c|l} \hline
			$Tom$ &	$C_{Tom}=\{K_{OA}, K_{AB}, C_{PCTom}, C_{PLCusr}, C_{IGSusr}\}		$ \\ \hline
			$Amy$ & $C_{Amy}=\{K_{OA}, K_{AB}, C_{PCAmy}, C_{IGSadm},C_{MBSLadm}\}		$ \\ \hline 
			$sup$ & $C_{sup}=\{K_{OA}, K_{AB}, C_{PCTom}, C_{PCAmy},$ \\ 	
			& $C_{PLCusr}, C_{IGSusr}, C_{IGSadm}, C_{MBSLadm}\} 					$ \\ \hline
		\end{tabular}
		\renewcommand{\arraystretch}{1.0}
	}
\end{table}
In order to have a better understanding of our approach, let us consider the simple INS sketched in Fig.~\ref{fig:cabinet}. The whole system is hosted in two rooms, $A$ and $B$, communicating through door $d_{AB}$. Room $O$ models a generic external environment.  Users in $O$ can enter room $A$ only if they own key $K_{OA}$ required to open the plant entrance door $d_{OA}$, while they can move from $A$ to $B$ (or from $B$ to $A$) by using key $K_{AB}$, needed to open door $d_{AB}$. Once in room $A$, a user can leave the plant area without using any credential. 

Room $B$ contains an industrial PC ($PLC$) running an ISaGRAF soft-PLC ($IGS$), an Ethernet  switch ($SW$) and a Modbus slave device ($MBSL$). $SW$ enables communications between $PLC$, $MBSL$ and other devices not located in $B$, such as the supervisory PC ($PC$) in room $A$. Dashed lines in Fig.~\ref{fig:cabinet} represent communication links.

To keep the example simple, we assume that only two roles, namely \textit{plant operator} ($P_o$) and \textit{plant supervisor} ($P_s$), have been defined by policy designers, and $P_o \prec P_s$, i.e., $P_s$ is higher than $P_o$ in the role hierarchy.
Moreover, users assigned to $P_o$ are enabled to perform operational activities ($run$) on both the ISaGRAF soft-PLC and Modbus slave, but they are not allowed to carry out management operations ($admin$) on the two objects. Conversely, users assigned to $P_s$ can administrate both the ISaGRAF soft-PLC and Modbus slave and are not explicitly  assigned denied permissions. Finally, only two users, $Tom$ and $Amy$, are assigned to roles $P_o$ and $P_s$, respectively. 
\begin{figure}[t!]
	\centering
	\includegraphics[width=0.85\linewidth]{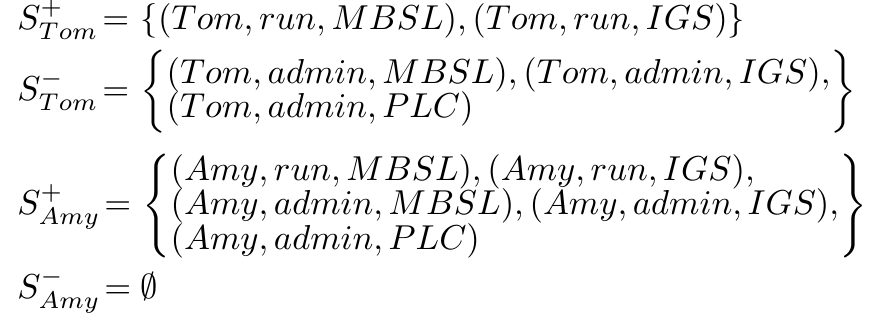}
	\caption{Specification sets $S^{\bullet}_{u}$, $\bullet=+, -$, for users $Tom$ and $Amy$.}
	\label{fig:u_tom_amy}
\end{figure}

The specification model $\mathcal{S}$ derived from the description above is summarized in Table~\ref{tab:policies}, where all roles are listed together with their explicitly allowed and denied permissions, as well as their user assignments. By using~(\ref{eq:SecSetDefUser}), and keeping in mind that $P_o \prec P_s$ so that~(\ref{p_asP}) and~(\ref{p_asM}) apply, we obtain the specification sets for $Tom$ and $Amy$ in Fig.~\ref{fig:u_tom_amy}.

\begin{figure*}[t!]
	\centering
	\framebox{
		\centering
		\includegraphics[width=1\linewidth]{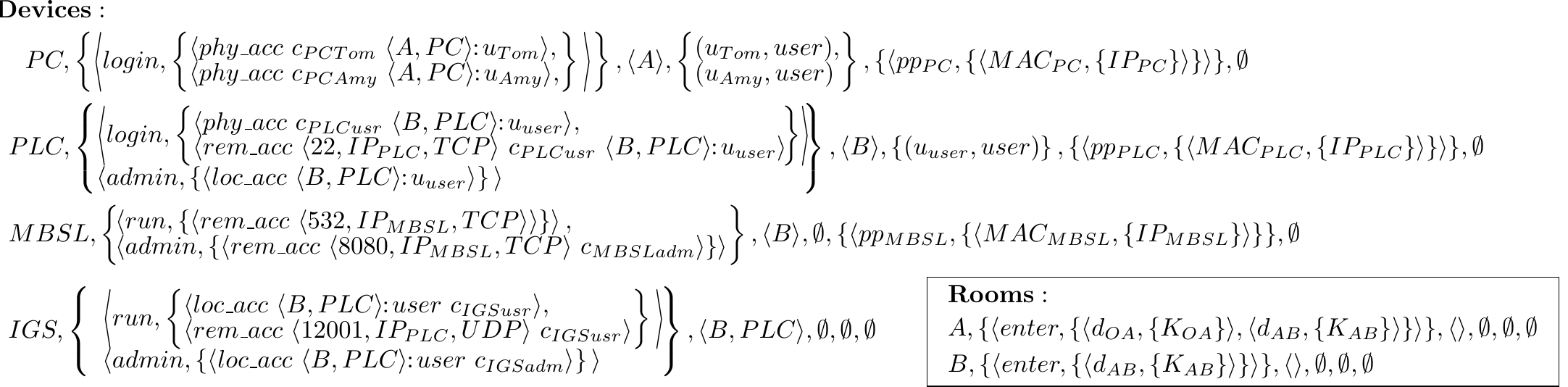}
	}
	\caption{Fragments of the data model}\label{fig:description}
\end{figure*}

Fig.~\ref{fig:description}, instead, shows those data elements in $\mathcal{I}$, that are mostly significant to understand the example, while the initial assignment of credentials to the two users is listed in Table.~\ref{tab:credential}.

In details object $PC$, contained in object $A$ (as highlighted by its path $\langle A \rangle$), is equipped with a single network port, $pp_{PC}$, bound to MAC address $MAC_{PC}$ and IP address $IP_{PC}$. A  $user$ group is defined for $PC$, which includes two accounts $u_{Tom}$ and $u_{Amy}$.  $PC$ supports only one operation ($login$), guarded by precondition $phy\_acc$. This means that a user shall have physical access to device $PC$ (i.e., s/he must be in room $A$) in order to log in. $login$ can be successfully performed by owning either credential (i.e., password) $C_{PCTom}$ or $C_{PCAmy}$, so that the user becomes logged on $PC$, respectively as $u_{Tom}$ or $u_{Amy}$, in the end. Note that the empty set at the end of the $PC$ description means that no filtering rule is defined for the device.

Similarly, object $PLC$ is contained in room $B$  and is equipped with one physical network port, $pp_{PLC}$, bound to MAC address $MAC_{PLC}$ and IP address $IP_{PLC}$. A $user$ group is defined on $PLC$ as well, consisting of a single account ($u_{user}$). Two operations are defined for $PLC$, namely $login$ and $admin$. A user can $login$ on $PLC$ in two alternative ways, that is by exploiting a physical access (i.e., the user is in room $B$) or remotely (precondition $rem\_acc$), through an SSH channel, if s/he is already logged on some host and a TCP connection exists to port $22$ of $PLC$ with IP address $IP_{PLC}$. In either cases the user needs to own credential $c_{PLCusr}$ which logs her/him on $PLC$ as $u_{user}$.

The $loc\_acc$ precondition for operation $admin$ means that the user must be logged on $PLC$ with any username belonging to group $user$ to be able to invoke it. This basically models the increase of privilege in accessing $PLC$, from simple user to administrator.

The Modbus slave, $MBSL$ supports two operations ($run$ and $admin$) that can be invoked remotely through TCP connections to ports $532$ and $8080$, respectively. No credential in needed to execute $run$, while $(admin, MBSL)$ can be performed only by owners of password $c_{MBSLadm}$.

$IGS$ models the ISaGRAF application running on $PLC$ as shown in its location path $\langle B,PLC\rangle$. Two operations, $run$ and $admin$, are defined also in this case. To ``run'' the soft-PLC, a user must be logged on either $PLC$ with a username belonging to group $user$ or  some other host connected to the $PLC$ port $12001$ via the UDP protocol. In both cases credential $c_{IGSuser}$ is necessary for authentication. Operation $admin$ can be executed by users logged on $PLC$ as members of group $user$ and owning credential $c_{IGSadm}$. Differently from $PC$ and $PLC$, no account is defined for $MBSL$ and $IGS$: only credentials are used to distinguish  between operational and administrative privileges. 

Room descriptions in the lower right corner of  Fig.~\ref{fig:description} assert that only operation $enter$ can be performed on $A$ and $B$ (through doors $d_{OA}$ and $d_{AB}$) by owning appropriate keys ($K_{OA}$ and/or $K_{AB}$). Note that the data model also includes  descriptions for the external environment $O$, switch $SW$ and communication links, but they are not shown here for conciseness reasons.	

\subsection{Policy verification and anomaly resolution}
\begin{figure}[t!]
 	\centering
 	\includegraphics[width=1\linewidth]{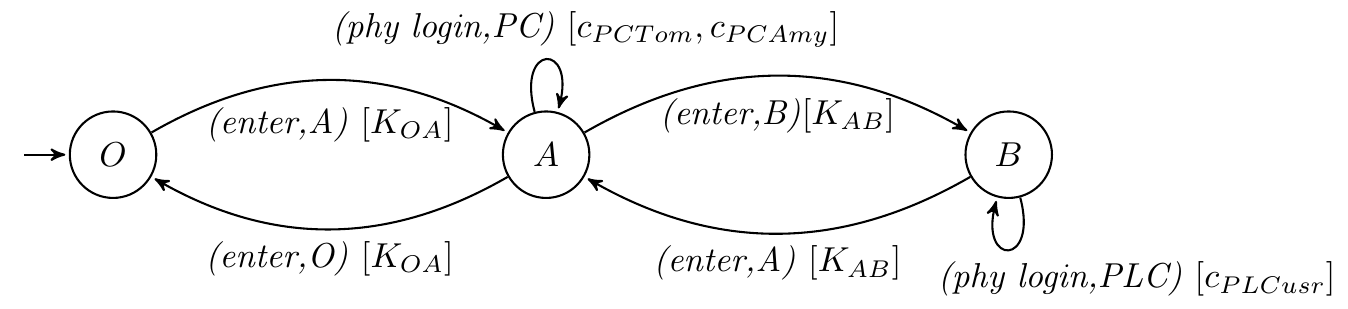}
 	\caption{Automaton $A^{r}$.}
 	\label{fig:ar}
\end{figure}

We now briefly describe the resulting automaton $A$ for the considered INS. We assume the super-user to be, initially, in the same room (i.e., in $O$) as $Tom$ and $Amy$. Fig.~\ref{fig:ar} shows $A^{r}$, describing the dynamics of the super-user owning all credentials defined in the system (see user $sup$ in Table.~\ref{tab:credential}), who moves among rooms. In the figure, labels in the form $(\pi, \omega)[c_1,c_2,\ldots,c_n]$ standing on single edge of $A^r$ represents $n$ different edges (all originating and ending in the same states as the original one) each one labeled as $(\pi,\omega,c_i)$. The automaton shows that the super-user can $enter$ room $A$ and $B$ in sequence (or move in the opposite direction), and, depending on whether s/he is in room $A$ or $B$, access $PC$ or $PLC$ (since the precondition $phy\_acc$ of the $login$ operation defined for both hosts is satisfied) by exhibiting the necessary credentials. 

\begin{figure}[t!]
	\centering
	\includegraphics[width=1\linewidth]{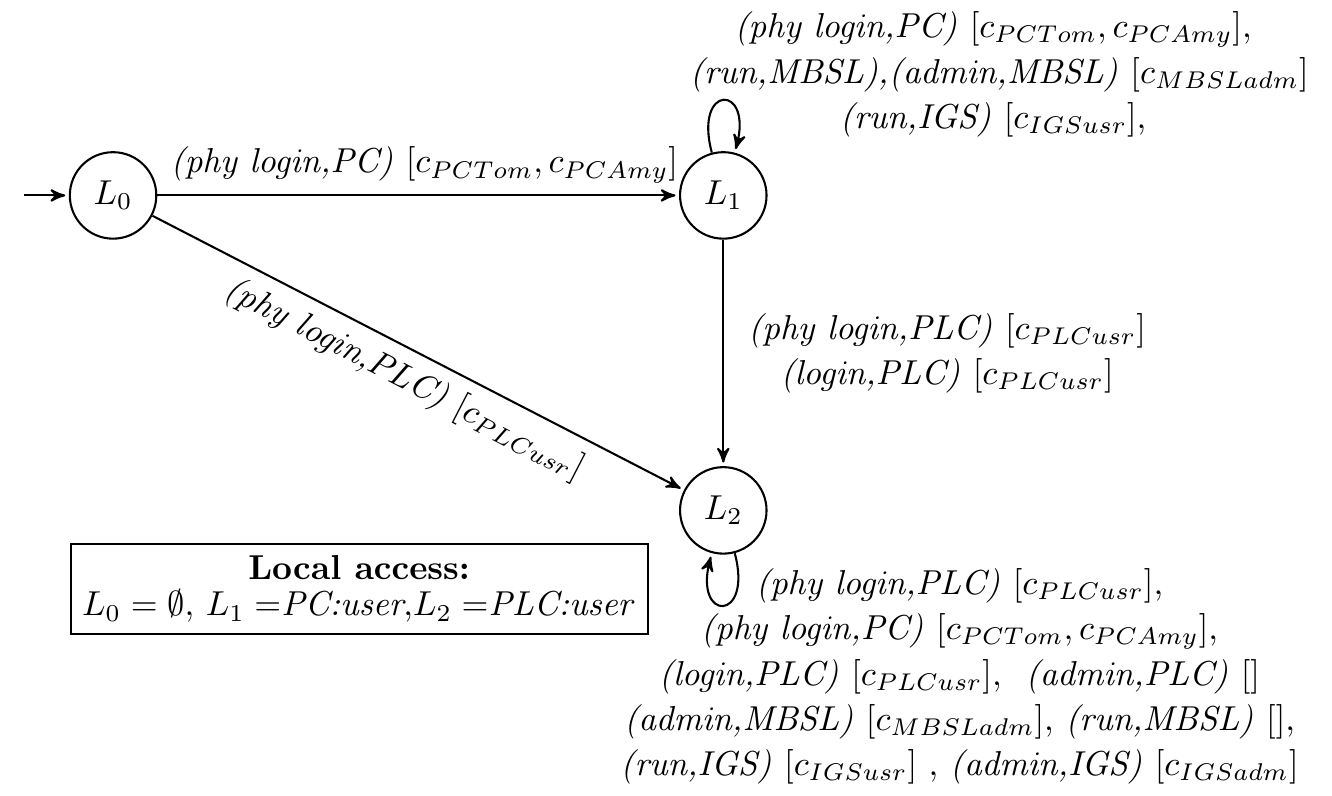}
	\caption{Automaton $A^{L,min}$.}
	\label{fig:al}
\end{figure}

The optimized local access automaton $A^{L,min}$ is shown in Fig.~\ref{fig:al}: language $\mathcal{L}(A^{L,min})$ describes the sequences of operations the super-user can perform, assuming all devices and their resources to be in her/his same virtual room  (i.e., it describes the dynamics of access to resources without considering their location). 

Starting from the initial state, the super-user can $login$ on either $PC$ or $PLC$ as s/he owns the required credentials (i.e., $c_{PCTom}$ or $c_{PCAmy}$ for $PC$ and $c_{PLCusr}$ for $PLC$).  The two actions ($(login,PC)$ and $(login,PLC)$) lead to different states in $A^{L,min}$, because the superuser logged on $PC$ can perform both $run$ and $admin$ on $MBSL$, but only run $IGS$ (by exploiting the remote connection),  since $admin$ on $IGS$ requires the user to be logged in on $PLC$. 

\begin{figure*}[t]
	\centering
	\includegraphics[width=1\linewidth]{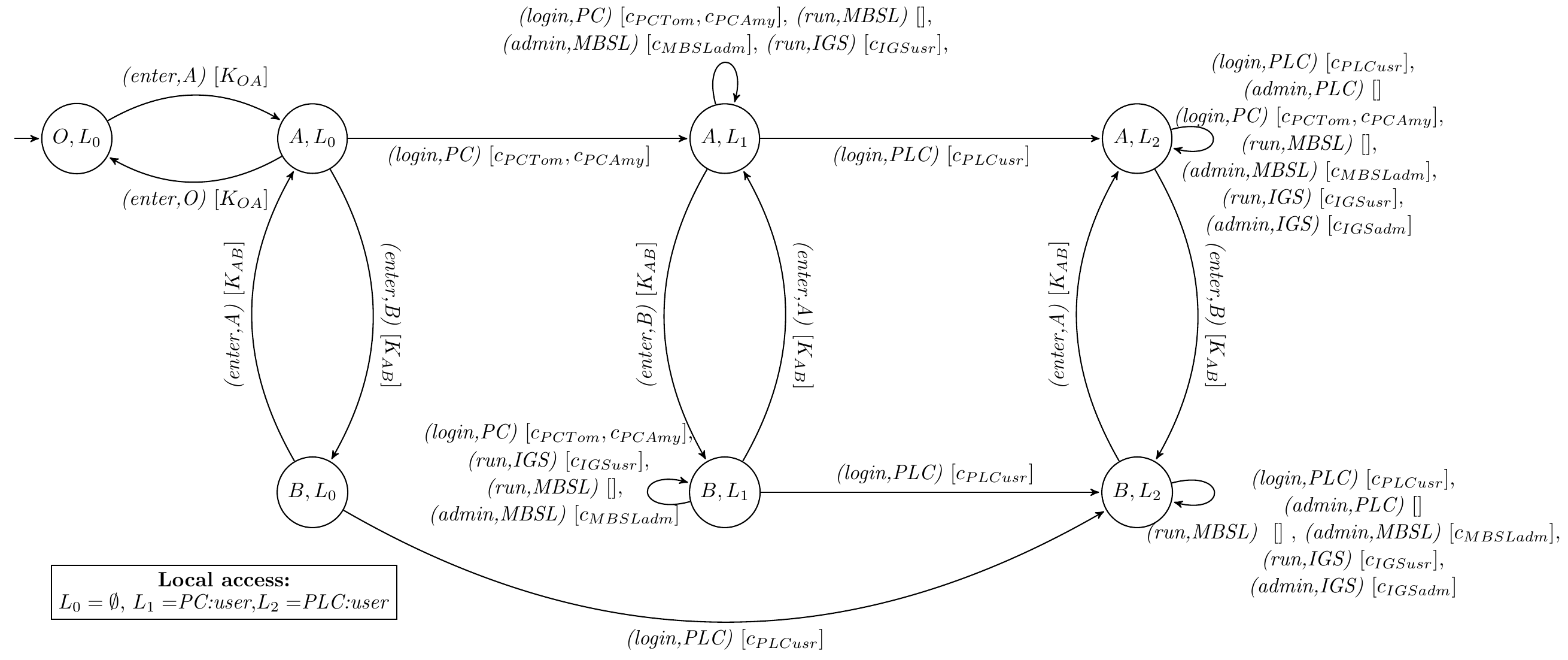}
	\caption{Automaton $A=A^{r}||A^{L,min}$.}
	\label{fig:a}
\end{figure*}

Automaton $A$ shown in Fig.~\ref{fig:a} is finally obtained from the parallel composition of $A^r$ and $A^{L,min}$ (and ignoring prefix $phy$ on transition labels which are only necessary for automata composition). 
From $A$, following the process described in Sect.~\ref{sec:hints_on_the_approach}, we derive the enabling functions of credentials $\mathcal{F}_u(e)$ shown in Fig.~\ref{tab:formule_max} for any permission $(\pi, \omega)$ appearing in the policies (i.e., in the specification sets $S^\bullet$). As an example, action $(enter, A)$ can only be performed by users owning credentials $K_{OA}$ (note that, actually $K_{OA}$ is necessary to perform any action in the system, and, as such the key appears in all functions in Fig.~\ref{tab:formule_max}). Analogously, action $(admin, IGS)$ requires a user to own credential $c_{IGSadm}$ and to either login on $PLC$ remotely, after logging in on $PC$ with credential $c_{PCTom}$ or $c_{PCAmy}$, or physically, meaning that the user needs key $K_{AB}$ and password $c_{PLCuser}$.

\begin{figure}[!h]
	\centering
	\includegraphics[width=\linewidth]{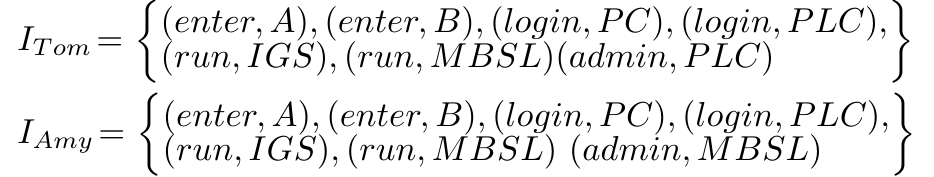}
	\caption{Implementation sets $I_{u}$ computed for users $Tom$ and $Amy$.}
	\label{fig:i_tom_amy}
\end{figure}

By computing enabling function values for users $Tom$ and $Amy$, assuming their credential sets are those described in Table.~\ref{tab:credential}, we derive the implementation sets shown in Fig.~\ref{fig:i_tom_amy}.

\begin{figure}[t!]
	\resizebox{1.05\columnwidth}{!}
	{
		\centering
		\renewcommand{\arraystretch}{1.3}
		\begin{tabular}{l} %\hline
			$\mathcal{F}((enter, A))=  K_{OA}$\\
			$\mathcal{F}((enter, B))=  K_{OA} \cdot K_{AB}$\\	
			$\mathcal{F}((log, PC))=  K_{OA}\cdot(c_{PCTom} + c_{PCAmy})$\\
			$\mathcal{F}((log, PLC))=  K_{OA}\cdot(c_{PCTom} + c_{PCAmy})\cdot c_{PLCusr}+ $\\
			\quad\qquad \qquad \qquad \quad $K_{OA}\cdot K_{AB}\cdot C_{PLCusr}$\\
			$\mathcal{F}((run, MBSL))=  K_{OA}\cdot(c_{PCTom} + C_{PCAmy})+ $\\
			\qquad \qquad \qquad \qquad \quad $ K_{OA}\cdot K_{AB} \cdot c_{PLCusr}$\\
			$\mathcal{F}((run, IGS))=  K_{OA}\cdot(c_{PCTom} + c_{PCAmy})\cdot c_{IGSusr}+ $\\
			\;\;\qquad \qquad \qquad \quad $ K_{OA}\cdot K_{AB}\cdot c_{PLCTom}\cdot c_{IGSusr}$\\			
			$\mathcal{F}((admin, PLC))= K_{OA} \cdot (c_{PCTom} + c_{PCAmy})\cdot c_{PLCusr}+ $\\
			\;\qquad \qquad \qquad \qquad \quad $ K_{OA}\cdot K_{AB}\cdot c_{PLCusr}$\\
			$\mathcal{F}((admin, MBSL))= K_{OA}\cdot(c_{PCTom} + c_{PCAmy})\cdot c_{MBSLadm}+$\\
			\;\quad\qquad \qquad \qquad \qquad \quad $ K_{OA}\cdot K_{AB}\cdot c_{PLCusr}\cdot c_{MBSLadm}$\\
			$\mathcal{F}((admin, IGS))=K_{OA}\cdot (c_{PCTom} + c_{PCAmy})\cdot c_{PLCusr}\cdot c_{IGSadm}+$\\
			\qquad \qquad \qquad \qquad \quad$ K_{OA}\cdot K_{AB}\cdot c_{PLCusr}\cdot c_{IGSadm}$\\ %\hline						
		\end{tabular}
		\renewcommand{\arraystretch}{1}
	}\\
	\caption{Enabling functions}\label{tab:formule_max}
\end{figure}

\begin{figure}[!h]
	\centering
	\includegraphics[width=1.\linewidth]{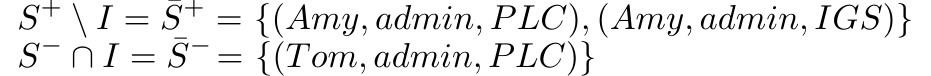}
	\caption{Conflicts highlighted by the analysis.}
	\label{fig:conflict}
\end{figure}

By applying (\ref{eq:Sp}), (\ref{eq:Sm}) and (\ref{eq:Spm0}) to sets $I=I_{Tom} \cup I_{Amy}$, $S^+=S^+_{Tom} \cup S^+_{Amy}$and $S^-=S^-_{Tom} \cup S^-_{Amy}$ (see Fig. \ref{fig:conflict}) we can see that the policies are not correctly implemented by the system as, differently from what expected, $Amy$ is allowed to administrate neither $PLC$ nor $IGS$, while $Tom$, that should not be allowed any administrator privilege is able to perform $(admin, PLC)$ in the current system implementation. 

To find user credential assignments that correctly enforce the specified access control policies (if any), we use a SAT solver to find solution(s) (if any) making (\ref{sol}) true. The result of the SAT solver analysis is shown in Table~\ref{tab:solution}, where the first rows show two possible correct credential assignments for user $Tom$ and the latter a single one for user $Amy$. As can be seen, to prevent $Tom$ from administrating $PLC$, he should be deprived of credential $c_{PLCusr}$, i.e., he should not be able to login on $PLC$ which is the action enabling $(admin, PLC)$. 
%Another option would be to define a required password to administrate $PLC$, which, anyhow is not always an available configuration option.
Moreover, the analysis shows that $Tom$ does not need to enter room $B$ (providing him with key $K_{AB}$ is unnecessary for the correct policy enforcement), as asserted by credential assignment $C^2_{Tom}$. Conversely, $Amy$ should be allowed to login on $PLC$ to perform both $(admin,PLC)$ and $(admin,IGS)$, as stated by the new credential assignment $C^1_{Amy}$ which includes password $c_{PLCAmy}$.

\begin{table}[h!]
	%\footnotesize
	\caption{Conflict resolution}\label{tab:solution}
	\scalebox{0.8}
	{
		\centering
		\renewcommand{\arraystretch}{1.3}
		\begin{tabular}{l| l} \hline
			\multirow{2}{*}{$Tom$} 	& $C^1_{Tom}=\{K_{OA}, c_{PCTom}, c_{IGSusr}\}$  \\ \cline{2-2}
			& $C^2_{Tom}=\{K_{OA}, K_{AB}, c_{PCTom}, c_{IGSusr}\}$ \\ \hline
			{$Amy$}  & $C^1_{Amy}=\{K_{OA}, K_{AB}, c_{PCAmy}, c_{PLCusr},c_{IGSadm},$\\ 
			&$ c_{MBSLadm}\}$  \\ \hline
		\end{tabular}
	}
	
\end{table}

\section{Conclusions and Future works}
\label{sec:conclusion_future_works}
This paper has presented an automated technique to fix access policy implementation anomalies in INSs. The proposed approach builds on previous works that dealt with the analysis of access policy through innovative twofold modeling techniques.  
 
The extended model and automated procedure we have described here enable: \textit{(i)} the high level definition of positive and negative access control policies and \textit{(ii)} the fine-grained description  of the industrial networked system implementation details. Discrepancies and errors found in the policy implementation can then be discovered by the verification process and possibly fixed by suitable (re)assignments of the user permissions. 

Future works will be aimed at extending this work in two directions: add different optimization strategies in order to automatically choose the best solutions for correction and increase the capability of our approach in order to also perform the refinement of high level policies to the INS low-level implementation.

%\section*{Acknowledgment}

%\bibliographystyle{IEEEtran}
%\bibliography{bibliography}


\end{document}